\documentclass[prd,twocolumn,floatfix,altaffilletter,preprintnumbers,tightenlines,showpacs,showkeys,nofootinbib,superscriptaddress]{revtex4-1}
\usepackage[colorlinks=true,citecolor=purple,linkcolor=purple,urlcolor=purple]{hyperref}
\usepackage[utf8]{inputenc}
\usepackage[normalem]{ulem}
\usepackage{amsmath,amssymb}
\usepackage{graphicx}
\usepackage{url}
\usepackage{color}
\usepackage[dvipsnames]{xcolor}
\usepackage{soul}
\usepackage[capitalise, english]{cleveref}
\usepackage{siunitx}
\AtBeginDocument{\usepackage{booktabs}}

\definecolor{tobycolour}{rgb}{.5,.0,.5}

\DeclareSIUnit\year{yr}
\DeclareSIUnit\pc{pc}

\allowdisplaybreaks

\bibpunct{[}{]}{,}{n}{}{,}

\graphicspath{{figs/}}

\begin{document}

\title{Ricci Reheating}

\author{Toby Opferkuch}
\email{toby.opferkuch@cern.ch}
\affiliation{Theoretical Physics Department, CERN, 1211 Geneva, Switzerland}
\author{Pedro Schwaller}
\email{pedro.schwaller@uni-mainz.de}
\author{Ben A. Stefanek}
\email{bstefan@uni-mainz.de}
\affiliation{PRISMA Cluster of Excellence and
             Mainz Institute for Theoretical Physics,
             Johannes Gutenberg-Universit\"{a}t Mainz, 55099 Mainz, Germany}

\date{\today}

\preprint{CERN-TH-2019-063, MITP/19-032}

\begin{abstract}
We present a model for viable gravitational reheating involving a scalar field directly coupled to the Ricci curvature scalar. Crucial to the model is a period of kination after inflation, which causes the Ricci scalar to change sign thus inducing a tachyonic effective mass $m^{2} \propto -H^2$ for the scalar field. The resulting tachyonic growth of the scalar field provides the energy for reheating, allowing for temperatures high enough for thermal leptogenesis. Additionally, the required period of kination necessarily leads to a blue-tilted primordial gravitational wave spectrum with the potential to be detected by future experiments. We find that for reheating temperatures $T_{\rm RH} \lesssim 1$ GeV, the possibility exists for the Higgs field to play the role of the scalar field.
\end{abstract}

\maketitle

\section{Introduction}
Inflation has established itself as a remarkably successful paradigm for avoiding fine-tuned initial conditions in cosmology.
There is growing evidence that the early universe underwent an accelerated period of expansion, likely driven by the potential energy of a slowly rolling homogeneous scalar field, the result of which was a very cold and flat universe~\cite{Akrami:2018odb}. It is much less clear, however, how reheating of the universe occurs after inflation. In particular, it is an open question as to whether successful reheating \emph{requires} the inflaton to have non-gravitational couplings to Standard Model (SM) fields. 

It is usually assumed that inflation ends when the inflaton rolls off a nearly flat region of its potential and begins to oscillate about some local minimum. If the inflaton couples to SM fields, its oscillations are damped and reheating occurs by decay of the inflaton field to SM particles. If however, the inflaton couples only gravitationally, it was realized that reheating could still potentially occur via the excitation of light, non-conformally coupled fields at the end of inflation~\cite{Ford:1986sy}. This leads to a gravitationally produced radiation energy density that is suppressed by $\sim 10^{-2} H_{*}^{2} /M_{P}^{2}$ relative to the dominant inflaton energy density, where $H_{*}$ is the Hubble rate at the end of inflation. In order for successful gravitational reheating to occur, the inflaton must have a post-inflationary equation of state $w$ which is stiffer than that of radiation ($w>1/3$) so that the gravitationally produced radiation can eventually come to dominate over the inflaton energy density and provide the correct initial condition for a hot Big Bang~\cite{Ford:1986sy,Spokoiny:1993kt}. A stiff equation of state is achieved when the kinetic energy of a scalar field dominates over its potential energy, leading to periods of cosmological history dominated by such a scalar to be called \emph{kination}~\cite{Ford:1986sy,Spokoiny:1993kt,Joyce:1996cp,Joyce:1997fc,Peebles:1998qn,Sahni:2001qp,Liddle:2003zw}.

It was recently pointed out that the requirement of a period of kination leads a blue-tilted primordial gravitational wave spectrum which dominates the energy density of the universe at late times. This leads to an inconsistent picture of gravitational reheating unless a large number of light fields were excited at the end of inflation~\cite{Figueroa:2018twl}.

In this work, we present a model of gravitational reheating involving at most one new degree of freedom. The model exploits the fact that the Ricci curvature scalar changes sign when the universe transitions from an inflationary quasi de-Sitter universe ($w\approx -1$) to a Friedmann-Robertson-Walker (FRW) universe dominated by an energy density with equation of state $w > 1/3$. The model consists of a non-minimally coupled scalar field $\phi$ which we call the \emph{reheaton}, which in general we take to be an extension of the SM matter content. However, we will also show that it is possible to identify the reheaton with the SM Higgs field, building upon the work of Ref.~\cite{Figueroa:2016dsc}. Due to its direct coupling to the Ricci scalar, the reheaton has an effective mass-squared which is positive $(m^{2} \propto H_{*}^2)$ during inflation but negative $(m^{2} \propto -H^2)$ after the universe transitions to kination domination. The effective tachyonic mass causes the reheaton field value to grow as it rolls toward the new minimum, thus both a non-zero vaccum expectation value (VEV) and energy density are achieved. Reheating occurs when the reheaton transfers its energy to SM radiation, which could happen before or after the field comes to dominate the energy density of the universe. We note that our model has a similar qualitative structure to that of Ref.~\cite{Dimopoulos:2018wfg}, but differs significantly in the quantitative details.

Successful reheating can be achieved for a wide region of the model parameter space and is relatively generic in the case where the energy density extracted by the reheaton scales as matter for some of the cosmological history. In the case of a long period of kination, we show that the resulting blue-tilted gravitational wave spectrum can be detected at future experiments for portions of the model parameter space. We find that the SM Higgs field can function as the reheaton, provided the reheating temperature is low enough to avoid large field excursions which would probe the unstable region of the Higgs potential.

This paper is organized as follows: After giving an overview of the model in \cref{sec:RROV}, we present the underlying details in \cref{sec:model,sec:DotR}. Reheating is discussed in \cref{sec:RH}, followed by prospects for detecting the blue-tilted primordial gravitational wave spectrum in \cref{sec:SGWB}. Observational constraints on the model are discussed in \cref{sec:constraints}. The parameter space where the SM Higgs field can be identified with the reheaton is shown in \cref{sec:HiggsRH}, before concluding in \cref{sec:conc}.

\section{Ricci Reheating: An Overview}
\label{sec:RROV}
%
\begin{figure}
	\includegraphics[width=\columnwidth]{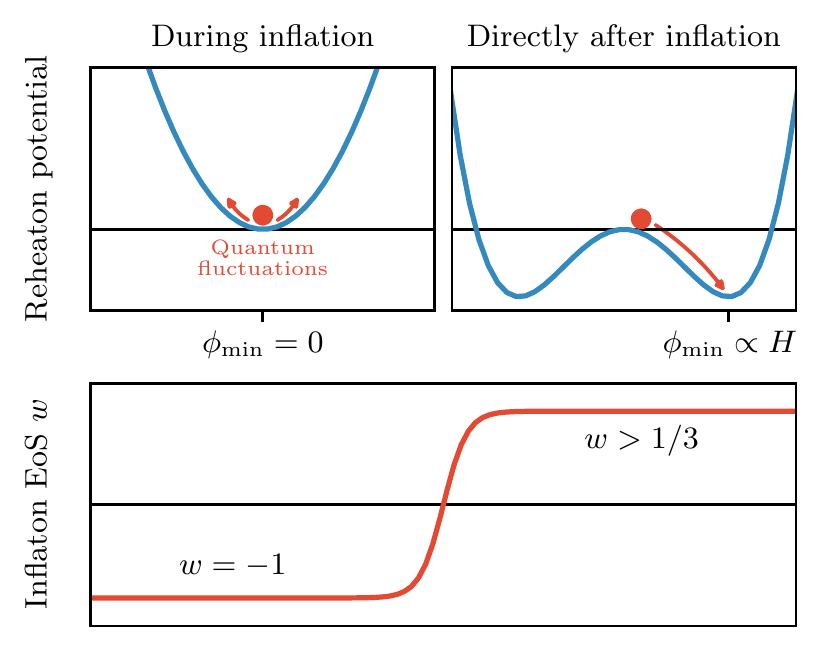}
	\caption{Dependence of the effective reheaton potential $V_{\rm eff}(\phi)$ on the inflaton equation of state (EoS) $w$. }
	\label{fig:sketch}
\end{figure}
If the inflationary sector has only gravitational couplings, the energy density of the inflaton at the end of inflation decays only due to the expansion of the universe as $\rho_{\rm inf} \propto a^{-3(1+w)}$, where $w$ is the post-inflationary equation of state and $a$ is the scale factor.  During inflation, the Ricci scalar $R$ is approximately constant with the value $R_{*} \approx -12H_{*}^{2}$, where stars denote quantities at the end of inflation. After inflation ends, $R$ in the post-inflationary FRW universe can be written as
\begin{equation}
R = -3(1-3w)H^{2}\,,
\label{eq:ricci}
\end{equation}
where $H = \dot{a}/a$ is the Hubble parameter. If the post-inflationary universe has an equation of state $w>1/3$, $R$ changes sign during the transition. Consequently, non-minimally coupled scalar fields, which have an effective potential of the form
\begin{equation}
V_{\rm eff}(\phi) = V(\phi)-\frac{1}{2}\xi R \phi^{2} \,,
\end{equation}
can become tachyonic for positive values of the non-minimal coupling $\xi$.

In what follows, we consider a model where the universe makes a transition from a quasi de-Sitter universe to an FRW universe dominated by the inflaton energy density described by an equation of state $w > 1/3$. This could be implemented by an inflationary potential which becomes steep at the end of inflation, such as large power monomials or non-oscillatory models~\cite{Ford:1986sy,PhysRevD.28.1243,Felder:1999pv,Spokoiny:1993kt,Peebles:1998qn,Sahni:2001qp,WITTEN1981267}. Beyond this and assuming a relatively fast transition, we will not specify the details of the inflationary sector, but rather simply treat the inflaton field as an order parameter that describes a transition in the equation of state of the universe as illustrated in \cref{fig:sketch}. 

Instead, we are mainly interested in the dynamics of the reheaton field $\phi$. 
For simplicity, we will focus on a bare potential of the form
\begin{equation}
V(\phi) = \frac{1}{2} m^{2} \phi^{2} +\frac{\lambda}{4}\phi^{4} \,,
\label{eq:vbare}
\end{equation}
where $m^{2}$ is constant. We also take $m^2, \lambda \phi_{*}^{2} \ll \xi H_{*}^2$ such that the non-minimal coupling dominates initially and there is a tachyon.
The steps in our mechanism can be summarized as:
\begin{itemize}
\item At the end of inflation, the inflaton does not decay but instead undergoes a change in its equation of state from $w \approx -1$ to $w > 1/3$.
\item This same transition causes the sign of the Ricci curvature scalar to change and the reheaton, a non-minimally coupled scalar field, to become tachyonic.
\item The tachyonic reheaton rolls toward the new minimum, acquiring a nonzero VEV $\langle \phi \rangle \propto H$ and energy density.
\item Tachyonic growth ceases when the reheaton is turned by the quartic, which happens shortly after it reaches the minimum of the effective potential.
\item Due to decreasing $H$, the $\xi R \phi^{2}$ term quickly becomes unimportant and the reheaton begins to effectively oscillate in its bare potential $V$. Depending on the size of $m$, the reheaton energy begins to scale as radiation ($V\approx \lambda \phi^{4}$) or matter ($V\approx m^{2}\phi^{2}$).
\item At some point, the reheaton transfers its energy to SM radiation. Reheating occurs when the SM radiation thermalizes and dominates the energy density of the universe.
\end{itemize}

\section{Model}
\label{sec:model}
We consider a theory in an FRW universe described by the metric 
\begin{align}
\label{eq:ds2}
ds^{2} =g_{\mu\nu}dx^{\mu}dx^{\nu} = dt^{2}  -  a^{2}(t)  \delta_{ij}dx^{i}dx^{j} \,,
\end{align}
with the following action
\begin{equation}
\mathcal{S} =\int d^{4}x \sqrt{-g} \left( -\frac{M_P^2}{2}R + \mathcal{L}_{\phi}  + \mathcal{L}_{\rm m}\right) \,.
\label{eq:act}
\end{equation}
The first term in $\mathcal{S}$ is the Einstein-Hilbert action with $M_P = 2.44\times 10^{18}\;{\rm GeV}$, $\mathcal{L}_{\phi}$ is the Lagrangian of the reheaton field $\phi$
\begin{equation}
\mathcal{L}_{\phi} = \frac{1}{2} g^{\mu\nu}\partial_{\mu}\phi\partial_{\nu}\phi +  \frac{1}{2}\xi R \phi^{2}- V(\phi) \,,
\end{equation}
and $\mathcal{L}_{\rm m}$ contains the remaining fields in the theory.\footnote{This includes the SM fields and the inflaton sector, as well as the couplings of $\phi$ to the SM.} The parameter $\xi$ controls the strength of the non-minimal coupling of $\phi$ to the Ricci scalar $R$, therefore $\xi = 0$ corresponds to minimal coupling.  The bare potential $V(\phi)$ given in \cref{eq:vbare} leads to an equation of motion for the reheaton of
\begin{equation}
\ddot{\phi}+ 3H\dot{\phi} + (m^{2}  - \xi R)\phi  +\lambda \phi^{3}= 0 \,.\,
\label{eq:eom}
\end{equation}
Varying $\mathcal{S}$ with respect to $g^{\mu\nu}$ yields the Einstein equations, which for the metric in \cref{eq:ds2} and a homogeneous field $\phi = \phi(t)$ reduce to the Friedmann equations
\begin{equation}
H^{2} = \left( \frac{\dot{a}}{a}\right)^{2} =\frac{\rho}{3M_{P}^{2}} \,,
\end{equation}
\begin{equation}
\dot{H} + H^{2} = \frac{\ddot{a}}{a} =  -\frac{1 }{6M_{P}^{2}}\left(\rho+ 3p\right) \,,
\end{equation}
where $\rho = \rho_{\rm m} + \rho_{\phi}$ and $p = p_{\rm m} + p_{\phi}$ are total energy density and pressure. In the case of a perfect fluid, the energy density and pressure are related by $p = w\rho$, which defines the equation of state $w$. At the end of inflaton, we assume that $\rho$ is still dominated by the inflaton energy density ($\rho \approx \rho_{\rm m} \approx \rho_{\rm inf} $), such that the Hubble rate scales as
\begin{equation}
H^{2}  \approx  \frac{\rho_{\rm inf}}{3M_{P}^{2}} \approx H_{*}^{2} \left(\frac{a_{*}}{a} \right)^{3(1+w)} \,.
\label{eq:H2inf}
\end{equation}
The reheaton energy density $\rho_{\phi}$, which we assume to be initially sub-dominant ($\rho_{\phi} \ll \rho_{\rm inf}$), is that of a homogeneous, non-minimally coupled scalar field~\cite{Ford:1987de,Ford:2000xg,PhysRevD.11.2072,PhysRevD.54.6233,DESER1984419,CALLAN197042,Bellucci:2001cc,Hrycyna:2015vvs}
\begin{equation}
\rho_{\phi} = \frac{1}{2}\dot{\phi}^{2} + V(\phi) + \xi\left(3H^{2}\phi^{2} + 6 H \phi \dot{\phi}\right)  \,.
\label{eq:rho_phi}
\end{equation}
The term multiplied by $\xi$, representing the new contribution to the energy density when the scalar field is non-minimally coupled, depends non-quadratically on $\phi$ and $\dot{\phi}$ and thus is not a priori guaranteed to be positive definite. That the energy density of a non-minimally coupled scalar can be negative has long been known and exploited in a variety of contexts~\cite{Ford:1987de,Ford:2000xg,PhysRevD.11.2072,PhysRevD.54.6233,Visser:1999de,Barcelo:1999hq,Barcelo:2000zf}. When working in an FRW universe, the situation is greatly simplified since the Einstein equations require $\rho = 3M_{P}^{2}H^{2}$ and thus the \emph{total} energy density in such a universe is positive at all times. This constraint, however, does not preclude the possibility of sub-components with negative energy. Further discussion on the energy-momentum of a non-minimally coupled scalar can be found in \cref{app:A}.

From the equation of motion given in \cref{eq:eom}, we see that $\phi$ can experience tachyonic growth if ${ m^{2} - \xi R <0}$. One could also interpret this in terms of an effective potential
\begin{equation}
V_{\rm eff}(\phi,R) = \frac{1}{2} (m^{2} - \xi R) \phi^{2} +\frac{\lambda}{4}\phi^{4} \,,
\label{eq:veff}
\end{equation}
which has a non-trivial minimum when the effective mass $m_{\rm eff}^{2} \equiv m^{2} - \xi R < 0$.  
Focusing on the case of positive $\xi$ and $m^{2} \ll \xi R$, we see from \cref{eq:ricci} that $m_{\rm eff}^{2} < 0$ requires $w > 1/3$ as previously discussed. In this case, $V_{\rm eff}$ has non-trivial minima at $\phi = \pm \phi_{\rm min}$ with
\begin{align}
	\phi_\text{min} = \left(\frac{\xi R}{\lambda}\right)^{\frac{1}{2}}= H\left(\frac{3\xi(3w-1)}{\lambda}\right)^{\frac{1}{2}}\,.
	\label{eq:phi_m}
\end{align}
Since they are set by the Hubble rate, the locations of the minima are time dependent. In an expanding universe, $a \propto t^{2/3(1+w)}$ grows with time and $H \propto a^{-3(1+w)/2}\propto t^{-1}$ decreases. Thus for $ 1/3 < w \leq 1$, the minima move toward the origin as $\phi_{\rm min} \propto a^{-2} - a^{-3}$. The depth of the minima are given by $V_{\rm eff}(\phi_{\rm min},R)$, which has the form
\begin{equation}
\Delta V_{\rm eff} = -\frac{\xi^{2}R^{2}}{4\lambda} = -\frac{9\xi^{2}}{4\lambda}(3w-1)^{2} H^{4} \,.
\end{equation}
We note that the depth of the minima depend very sensitively on the Hubble rate as $\Delta V_{\rm eff}\propto H^{4}$. In an FRW universe with $1/3 < w \leq 1$ this translates into $\Delta V_{\rm eff} \propto a^{-8} -a^{-12}$.

\section{Dynamics of the Reheaton}
\label{sec:DotR}
After inflation ends, the effective reheaton mass becomes tachyonic and the reheaton is displaced from the unstable maximum at the origin by quantum fluctuations. It then rolls to the new minimum in approximately an e-fold, before proceeding to be turned by the quartic. On the return journey, the reheaton is not trapped in the tachyonic minimum since its depth is decreasing in time as $H^4$. Instead, the reheaton rolls over the bump at the origin and proceeds to the other side of the potential. As the tachyon continues to diminish in importance, the reheaton quickly begins to oscillate in the bare potential given by \cref{eq:vbare}. We now describe each of these phases in more detail, and estimate the energy extracted by the reheaton.

\subsection{Initial Dynamics}
\label{sec:init_dyn}
We take the initial conditions for the reheaton to be set by the root mean square of its quantum fluctuations at the end of inflation, which for $\xi \gtrsim 0.1$ we calculate to be
\begin{equation}
\phi_{*}  \approx \frac{H_{*}}{2\pi} \sqrt{\frac{H_{*}}{3m_{*}}} \,, \hspace{12.5mm} \, \dot{\phi}_{*} \approx m_{*} \phi_{*} \,,
\label{eq:phi_IC}
\end{equation}
where $m_{*} \approx \sqrt{-\xi R_{*}} \approx \sqrt{12\xi} H_{*}$ is the effective mass of the reheaton at the end of inflaton. The details of the computation can be found in \cref{app:B}. Our approach here is to take \cref{eq:phi_IC} as a conservative estimate of the size of the pre-tachyonic fluctuations and model the tachyonic growth after $R$ changes sign classically. We see that for $\lambda \lesssim 1$, we have $\phi_{*} \ll \phi_{\rm min}(t_*)$ and thus the initial rolling of $\phi$ is governed by the tachyonic mass $m_{\rm eff}^{2} \approx -\xi R$. Therefore, until $\phi$ reaches the minimum, it obeys
\begin{equation}
\ddot{\phi}+ 3H\dot{\phi} - \xi R\phi \approx 0 \,,
\end{equation}
which has a growing solution of the form
\begin{equation}
\phi(t) \approx \frac{\phi_{*}}{2} \left(1+\frac{\gamma_{*}}{\gamma}\right) \left(\frac{H_{*}}{H} \right)^{\gamma} \,,
\label{eq:phi_scal}
\end{equation}
with the exponent defined as $\gamma \equiv \sqrt{\xi(3w-1)/3}$ and $\gamma_{*}  \approx \sqrt{12\xi}/3$. Since it grows as $H^{-\gamma}$, the reheaton rolls to the minimum faster than the minimum approaches the origin only if $\gamma > 1$, which requires $\xi > (w-1/3)^{-1}$. This condition has implications for the initial scaling of the reheaton energy density which are discussed in \cref{app:A}.

\subsection{Time to the Minimum}
\label{section:TttM}
Since $H^{-1}$ is the only relevant timescale, it will prove useful to express the time that the reheaton takes to reach the minimum in terms of the number of e-folds after the end of inflation $N = \int H dt = \log (a /a_*)$
\begin{equation}
N_{\rm min} = \log\left(\frac{a_{\rm min}}{a_{*}} \right) = \frac{2}{3(1+w)}\log\left(\frac{H_{*}}{H_{\rm min}}\right) \,,
\end{equation}
where we have used \cref{eq:H2inf} to express the logarithm in terms of $H$.
The Hubble rate $H_{\rm min}$ when the reheaton reaches the minimum is found by comparing \cref{eq:phi_m} and \cref{eq:phi_scal}
\begin{equation}
\left(\frac{H_{\rm min}}{H_{*}} \right)^{1+\gamma} = \frac{1}{36\pi}\sqrt{\frac{\lambda}{\gamma_{*}}} \frac{(\gamma_{*} + \gamma)}{\gamma^{2}} \,.
\end{equation}
%
Parametrically, we have $(H_{\rm min}/H_{*})^{1+\gamma} \propto \sqrt{\lambda /\xi^{3/2}}$. To get an idea of how much the Hubble rate has changed when the reheaton reaches the minimum, we note that for $\xi, w \sim \mathcal{O}(1)$, we find $H_{\rm min} \sim 0.1 \, \lambda^{1/4}H_{*}$ which evaluates to $H_{\rm min} \sim 10^{-2} \, H_{*}$ for typical values of $\lambda$, namely $10^{-5} \lesssim \lambda \lesssim 10^{-2}$. The typical timescale to reach the minimum is thus $N_{\rm min} \sim \mathcal{O}(1)$, by which time the Hubble rate has generally decreased by a factor of $\mathcal{O}(10^{2})$. 
\subsection{Dynamics after the Minimum}
From \cref{sec:model} and \cref{section:TttM}, we find that the reheaton typically reaches the minimum within $\mathcal{O}(1)$ e-fold, but that by this time, the location of the minimum has moved closer to the origin by a factor of $\sim 10^{2}$ and the depth of the minimum has decreased by a factor $\sim(10^{2})^{4} = 10^{8}$. This means that after the reheaton proceeds past the minimum and is turned by the quartic, it has enough energy upon its return to go over the bump at the origin and proceed to the other side of the potential. Thus, after reaching the minimum, the reheaton quickly approaches the dynamics of a scalar field oscillating in a quartic potential. This implies that $\langle \phi \rangle = 0$ and that $\phi$ has an averaged equation of state $\langle w_{\phi} \rangle = 1/3$, so the energy density of $\phi$ scales like radiation.

Radiation scaling of the reheaton energy continues until it transfers its energy to the SM, or the amplitude of the field damps down enough such that the bare mass term becomes important. In the latter case, the energy density in $\phi$ begins to scale like matter. The field value $\phi_{\rm matt}^{2} = 2m^{2} /\lambda$ where this happens can be found by setting the two components of \cref{eq:vbare} equal. While oscillating in the quartic, the amplitude of $\phi$ decays as $\phi \propto a^{-1} \propto e^{-N}$, so we can use  $\phi_{\rm min}$ as a reference point and estimate the time $N_{\rm matt} = N_{\rm min} + \log \left(\phi_{\rm min}/\phi_{\rm matt}\right) $ when matter scaling takes over. Using \cref{eq:phi_m}, we find
\begin{equation}
N_{\rm matt} = N_{\rm min} + \log \left(\frac{3\gamma }{\sqrt{2}} \frac{H_{\rm min}}{m} \right) \,.
\label{eq:Nmatt}
\end{equation}
We support the statements of this section by numerically solving the equations of \cref{sec:model} in \cref{app:NMC_dyn}. 

\subsection{Energy Extracted by the Reheaton}
In the previous section we argued that the reheaton energy density begins to scale like radiation after being turned by the quartic for the first time. Thus, we would like to estimate the energy density $\rho_{\phi}^{1}$ at this time, since we know the scaling afterwards. At the turning point, all derivatives of $\phi$ vanish, so the energy density is given by \cref{eq:rho_phi} with $\dot{\phi} = 0$. Neglecting the bare mass term, we have
\begin{equation}
\rho_{\phi}^{1} = \frac{\lambda}{4}\phi_{1}^{4} +3\xi H_{1}^{2} \phi^{2}_{1}\,.
\end{equation}
Since the first turning point occurs shortly after the reheaton passes through the minimum, we make the approximation $N_{1} \approx N_{\rm min}$ and $\phi_{1} \approx \phi_{\rm min}$ (which we will also justify numerically) to approximate the energy density at the first turning point as $\rho_{\phi}^{1} \approx \Lambda$, with
\begin{equation}
\Lambda = (1+w)(3w-1) \frac{27\xi^{2}}{4\lambda}H_{\rm min}^{4} \,,
\label{eq:Lambda}
\end{equation}
where we have used \cref{eq:phi_m}.
The quantity $\Lambda$ given in \cref{eq:Lambda} represents the energy extracted via our mechanism, which depends on the model parameters as $\Lambda \propto \xi^{2} H_{\rm min}^{4} / \lambda$. 
\begin{figure}
	\includegraphics[width=\columnwidth]{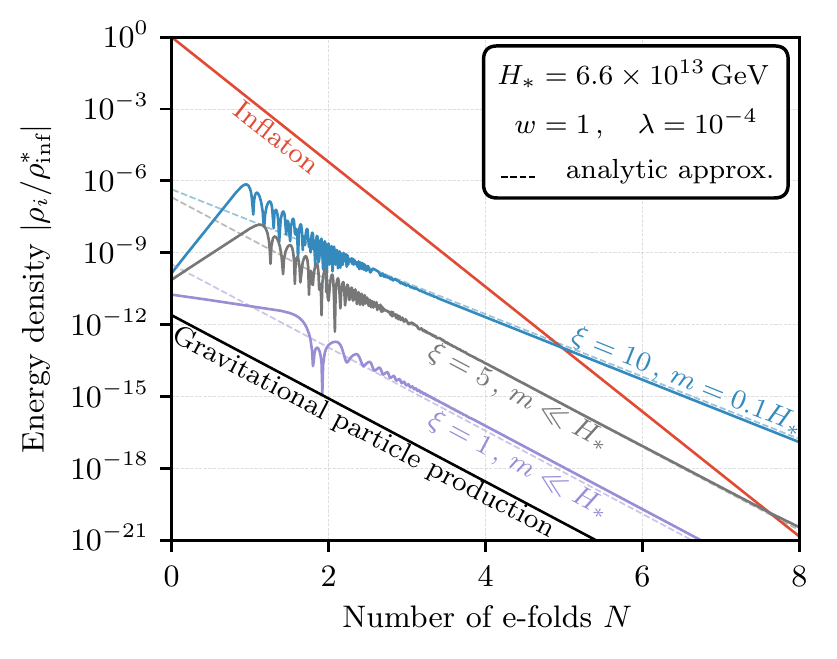}
	\caption{Reheaton energy density (normalized to the initial inflaton energy density $\rho_{\rm inf}^{*} \approx 3M_P^{2} H_{*}^{2}$) as computed numerically (solid lines), and compared to the analytic effective energy density approximation (dashed lines) given in \cref{rho_phi_eff}, for several values of the model parameters. Also shown for comparison is the inflaton energy density and the energy density expected from gravitational particle production, $\rho_{\rm GPP} \approx 10^{-2} H_*^4$.}
	\label{fig:phi_energy_density}
\end{figure}
We now define an effective energy density using the fact that $\Lambda$ dilutes as radiation until the bare mass becomes important at $N = N_{\rm matt}$ and matter scaling begins. The general form can be summarized as
\begin{equation}
	\rho_{\phi}^{\rm eff}(N) = \Lambda \, e^{-4(N-N_{\rm min})} \begin{cases}
		 1 \,, & N < N_{\rm matt} \\ 
		e^{N-N_{\rm matt}}  \,, & N \geq N_{\rm matt}
			\end{cases}\,.
\label{rho_phi_eff}
\end{equation}
In \cref{fig:phi_energy_density} we show the effective reheaton energy density defined in \cref{rho_phi_eff} compared to the reheaton energy density computed from our numerical solution of the system for several values of the model parameters. The numerical solution shows an initial tachyonic growth period as the reheaton rolls to the minimum, a decaying transition region where the reheaton energy density can become dominated by the $H\phi \dot{\phi}$ term and thus oscillate negative, and finally either radiation or matter asymptotic scaling as the reheaton begins to effectively oscillate in its bare potential. We see that the agreement between the numerical solution the effective energy density is quite good in the asymptotic regime. Since this is the regime where reheating occurs, we will use $\rho_{\phi}^{\rm eff}$ as a tool for analytically determining the reheating temperature in the next section. For the interested reader, the pre-asymptotic regime is discussed further in \cref{app:A}.
 
\section{Reheating}
\label{sec:RH}
In this section we estimate the reheating temperature by assuming that some process with rate $\Gamma$ is responsible for transferring the energy from the reheaton to SM radiation. In an expanding universe, a process with rate $\Gamma$ becomes efficient when $\Gamma \sim H$, at which time we will assume a fast transfer of energy.
For simplicity, we will also assume that $\Gamma < H_{\rm min}$ so that the reheaton always reaches the minimum before the energy transfer and we can apply the effective energy density approach of the previous section. This criterion, however, is not strictly necessary and one may entertain the possibility of a reheaton with large couplings to the SM which transfers its energy before reaching the minimum. 

For the purpose of defining the reheating temperature, we consider successful reheating to occur when i) the reheation field has transferred its energy to SM radiation, ii) the SM radiation has reached thermal equilibrium, and iii) the SM radiation bath dominates the energy density of the universe. We acknowledge the possibility that the last assumption may be relaxed, but do not study it here.

\subsection{Reheating after Reheaton-Kination Equality}
Kination ends when the energy density of the reheaton equals that of the inflaton. We denote the Hubble rate at this point as $H_{\rm ke}$, and reheating after this point means we should have $\Gamma < H_{\rm ke}$. Because $\phi$ is assumed to dominate the energy density, it has the critical energy density $\rho_{\phi} = 3M_{P}^{2}H^{2}$ and we can compute the reheating temperature simply by assuming that the reheaton energy is converted into SM radiation when $H\sim\Gamma$ which then quickly thermalizes
\begin{equation}
\rho_{\phi} = 3M_{P}^{2}H^{2} \bigg|_{H = \Gamma}= \frac{\pi^{2}}{30}g_{\rm RH} T_{\rm RH}^{4} \,,
\end{equation}
\begin{equation}
 T_{\rm RH} = \left( \frac{90}{\pi^{2} g_{\rm RH}}\right)^{1/4} \sqrt{\Gamma M_P} \,,
\end{equation}
where $g_{\rm RH}$ is the number of relativistic SM degrees of freedom at $T_{\rm RH}$. We see that this case reduces to the standard formula for $T_{\rm RH}$, e.g. from inflaton decay, where the reheating temperature is completely determined by the microphysics represented by $\Gamma$.

\subsection{Reheating at Reheaton-Kination Equality}
\label{sec:mainRH}
Here we assume that the reheaton transferres its energy to SM radiation while the inflaton still dominates the energy density. Reheating is then defined by the point where the energy in the SM radiation bath equals the energy of the inflaton, such that the SM bath has half the critical density
\begin{equation}
\rho_{\rm SM} = \frac{3}{2}M_{P}^{2}H^{2} \bigg|_{H = H_{\rm ke}}= \frac{\pi^{2}}{30}g_{\rm RH} T_{\rm RH}^{4} \,,
\end{equation}
and we find the reheating temperature to be
\begin{equation}
T_{\rm RH}= \left( \frac{45}{\pi^{2}g_{\rm RH} } \right)^{\frac{1}{4}} \sqrt{M_{P} H_{\rm ke}}\,.
\label{eq:TrhMAIN}
\end{equation}
Here the Hubble rate at reheaton-kination equality $H_{\rm ke}$, derived in \cref{app:GW}, is given as 
\begin{equation}
	\frac{H_{\rm ke}}{H_{*} } = \sqrt{2} \, \Omega_{\phi,*}^{\frac{3(1+w)}{2(3w-1)}} \begin{cases}
		 1 \,, & \Gamma > H_{\rm matt} \\ 
		\left(\frac{H_{\rm matt}}{\Gamma} \right)^{\frac{1}{3w-1}} \,, & \Gamma < H_{\rm matt}
			\end{cases}\,,
\label{eq:Hke}
\end{equation}
with $H_{\rm matt} = H_{*} \exp[-3(1+w)N_{\rm matt} /2]$.
The parameter $\Omega_{\phi,*}$ is the initial energy density fraction of the reheaton using the effective energy density approach of \cref{rho_phi_eff}
\begin{equation}
\Omega_{\phi,*} = \frac{\rho_{\phi}^{\rm eff}(N=0)}{3M_{P}^{2}H_{*}^{2}} = \frac{\Lambda\, e^{4N_{\rm min}}}{3M_{P}^{2}H_{*}^{2}} \,.
\end{equation}
The two cases in \cref{eq:Hke} correspond to whether the energy in the reheaton is transferred to SM radiation before ($\Gamma > H_{\rm matt}$) or after the reheaton amplitude has damped down enough for the $m^{2} \phi^{2}$ term to become important and the reheaton energy begins scaling like matter.
In the latter case  ($\Gamma < H_{\rm matt}$), the extra factor in \cref{eq:Hke} accounts for the fact that the reheaton energy density is larger compared to radiation only scaling by a factor $a_{\Gamma}/a_{\rm matt}$, where $a_{\rm \Gamma}$ is the scale factor at the time of transfer. 

\begin{figure}
	\includegraphics[width=\columnwidth]{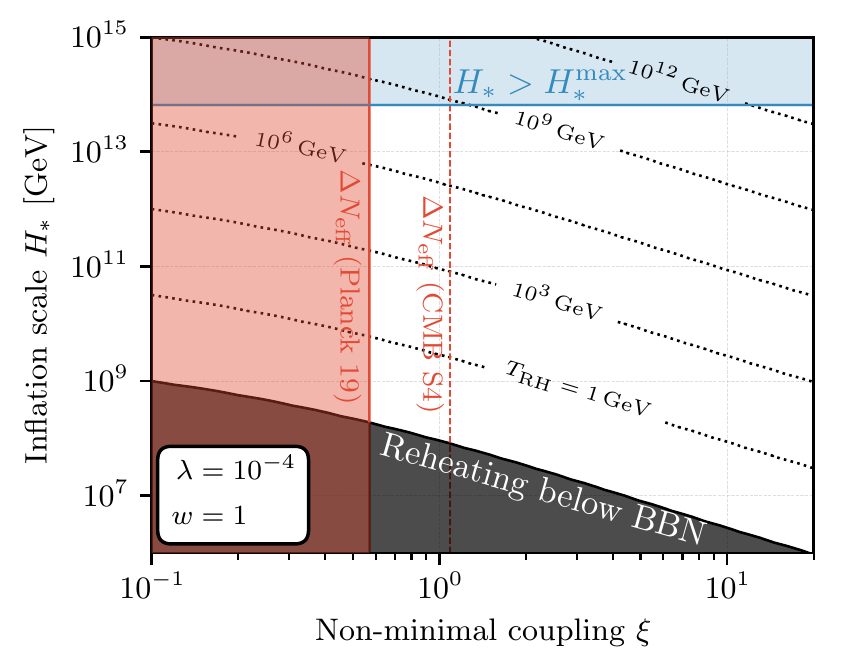}
	\caption{
	Values of the non-minimal coupling $\xi$ and inflation scales $H_*$ for which successful reheating is possible, for $\lambda = 10^{-4}$ and $w=1$. The upper bound on $H_*$ comes from the non-observation of B-modes in the CMB~\cite{Akrami:2018odb}, while the $\Delta N_{\rm eff}$ constraint is explained in \cref{sec:Neff}. 
}
	\label{fig:genericRH}
\end{figure}

The $\Gamma > H_{\rm matt}$ case is particularly interesting for us, since the reheating teperature depends only on the reheaton dynamics and not on $\Gamma$. In \cref{fig:genericRH} we show contours of constant reheating temperature for a $\lambda = 10^{-4}$ slice of the parameter space, as well as the regions excluded by the upper bound on the inflation scale, reheating below BBN, and overproduction of gravitational waves (see \cref{sec:Neff}). 
	
The reheating temperature for a fixed choice of $\xi$ can be increased by choosing a higher inflation scale $H_*$ or by reducing the size of $\lambda$. For $\xi = w = 1$, the reheating temperature scales roughly as
\begin{equation}
T_{\rm RH} \approx 10^{3} \,\text{GeV} \left( \frac{H_{*}}{10^{11}\,\text{GeV}}\right)^{2} \left( \frac{\lambda}{10^{-4}} \right)^{-1/5} \,,
\end{equation}
so the dependence on $\lambda$ is relatively weak. This remains true even for larger values of $\xi$, such as $\xi = 4\pi$ where $T_{\rm RH} \propto \lambda^{-1/2}$. For smaller values of $w$, the reheating temperature is lower for fixed parameters and $T_{\rm RH}$ depends more sensitively on $H_{*}$ and $\lambda$.
The reheating temperature cannot be arbitrarily increased by decreasing $\lambda$, since as $\lambda$ becomes increasingly small the reheaton field will eventually undergo trans-Planckian field excursions. For $\xi=4\pi$, $w=1$ and $H_ * = \SI{6.6E+13}{\GeV}$ this occurs for $\lambda \lesssim \SI{3E-12}{}$, leading to a reheating temperature of $\SI{6E+15}{\GeV}$ for $\phi_\text{min} = M_P$. 

Finally, we note that the reheating temperatures are generically higher in the case where the reheaton had a period of matter scaling, as its energy density would not have been diluted as much as the pure radiation case. However, in both cases it is possible to achieve reheating temperatures which are sufficiently high for successful thermal leptogenesis, namely $T_{\rm RH} \gtrsim 10^{8}$ GeV.

\section{Stochastic Gravitational Wave Background}
\label{sec:SGWB}
In the case of a period of kination, the primordial gravitational wave spectrum from inflation becomes blue-tilted~\cite{Giovannini:1998bp,Giovannini:1999hx,Giovannini:1999bh,Giovannini:2008zg,Giovannini:2008tm,Peebles:1998qn,Sahni:2001qp,Li:2016mmc,Caprini:2018mtu,Riazuelo:2000fc,Tashiro:2003qp,Boyle:2007zx,Artymowski:2017pua}. Long periods of kination can result in GW amplitudes which are large enough to be detectable.  However, excessive blue-tilting may lead to an unacceptable contribution of the gravitational wave energy density to $N_{\rm eff}$ (see \cref{sec:Neff}). 

The energy density fraction of primordial gravitational waves upon re-entering the horizon scales as
\begin{equation}
\Omega_{\rm GW}(k) = \frac{1}{\rho} \frac{d\rho_{\rm GW}}{d\log k} \propto a^{3w-1} \,,
\end{equation}
which is a growing function of $a$ for $w>1/3$. Because horizon crossing for a given mode is defined by $k = aH$, using \cref{eq:H2inf} we can derive the tilt of the spectrum
\begin{equation}
\Omega_{\rm GW}(k) \propto \left( \frac{k}{k_{\rm ke}}\right)^{\frac{2(3w-1)}{1+3w}} \,,
\end{equation}
which should be applied to all modes $k \geq k_{\rm ke}$ where $k_{\rm ke} = a_{\rm ke} H_{\rm ke}$ is the comoving momentum of a mode crossing the horizon at reheaton-kination equality.
\begin{figure}
	\includegraphics[width=\columnwidth]{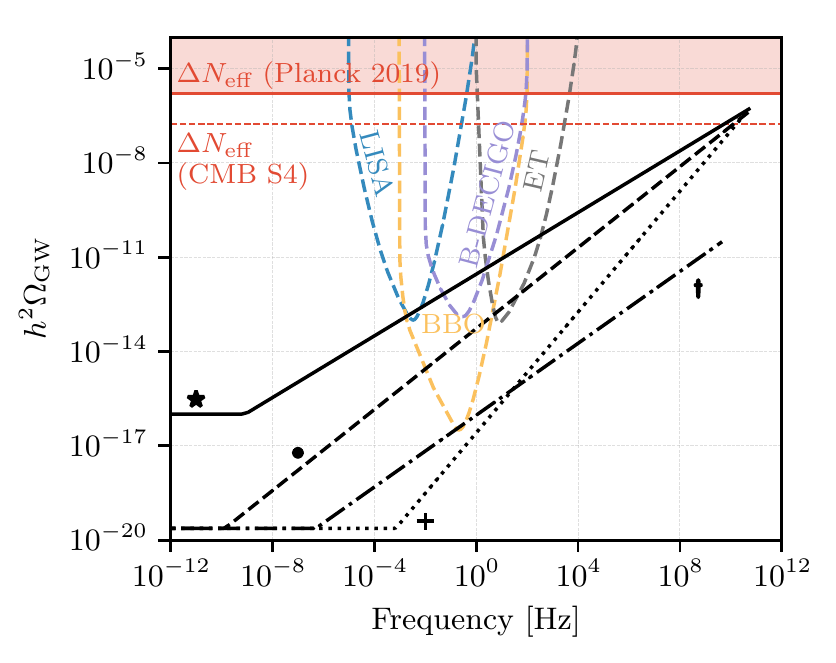}
	\caption{Gravitational wave spectra for the benchmark points of \cref{tab:BMPs} as well as the power-law integrated sensitivity curves of upcoming experiments.}
	\label{fig:GWspec}
\end{figure}
 Assuming the reheaton transferred its energy to SM radiation before kination ended, the physical momentum today of this scale is
\begin{equation}
k_{\rm ke}^{0} = H_{\rm ke}\left(\frac{a_{\rm ke}}{a_{0}} \right) = H_{\rm ke}\left(\frac{g_{s, \rm eq} }{g_{s,\rm ke}} \right)^{\frac{1}{3}}\left(\frac{T_{0}}{T_{\rm ke}} \right) \,,
\label{eq:kke0}
\end{equation}
where $g_{s, \rm eq} = 2+2N_{\rm eff}(7/8)(4/11) = 3.938$, $T_{0} = 2.35\times 10^{-13}$ GeV is the photon temperature today, and $T_{\rm ke}$ is the temperature of the SM radiation at the end of kination.  To produce an observable GW signal, a long period of kination is required, which is most easily achieved in the case where the reheaton energy never scales like matter. Focusing on this case, we have $T_{\rm ke} = T_{\rm RH}$ with $T_{\rm RH}$ given by \cref{eq:TrhMAIN} in the $\Gamma > H_{\rm matt}$ case. The spectrum has a UV cutoff given by the size of the horizon at the end of inflation, redshifted to the present time
\begin{equation}
k_{*}^{0} = H_{*} \left(\frac{g_{s, \rm eq} }{g_{s,\rm ke}} \right)^{\frac{1}{3}} \left(\frac{T_{0}}{T_{\rm ke}} \right) \left(\frac{H_{\rm ke}^{2}}{2H_{*}^{2}} \right)^{\frac{1}{3(1+w)}} \,,
\end{equation}
a derivation of which can be found in \cref{app:GW}.
\begin{table}
	\begin{tabular}{l c c c c } 
	\toprule
Benchmark point & $w$ & $H_*$ [GeV] & $\xi$ & $\lambda$  \\\midrule
	1. LISA `$\boldsymbol{\star}$' & $0.55$ & $\SI{6.6E+13}{}$ & $1.0$ & $\SI{E-4}{}$ \\ 
	2. ET `$\bullet$' & $0.65$ & $\SI{E+12}{}$ & $1.0$ & $\SI{E-4}{}$ \\ 
	3. CMB-S4 `$\boldsymbol{+}$' & $0.95$ & $\SI{E+12}{}$ & $0.8$ & $\SI{E-4}{}$  \\ 
	4. Large $\xi$ `$\boldsymbol{\dagger}$' & $0.6$ & $\SI{E+12}{}$ & $15.0$ & $\SI{E-4}{}$\\ 
	\bottomrule
	\end{tabular}
	\caption{Benchmark point model parameters.}
	\label{tab:BMPs}
\end{table}
\begin{figure*}
	\includegraphics[width=\textwidth]{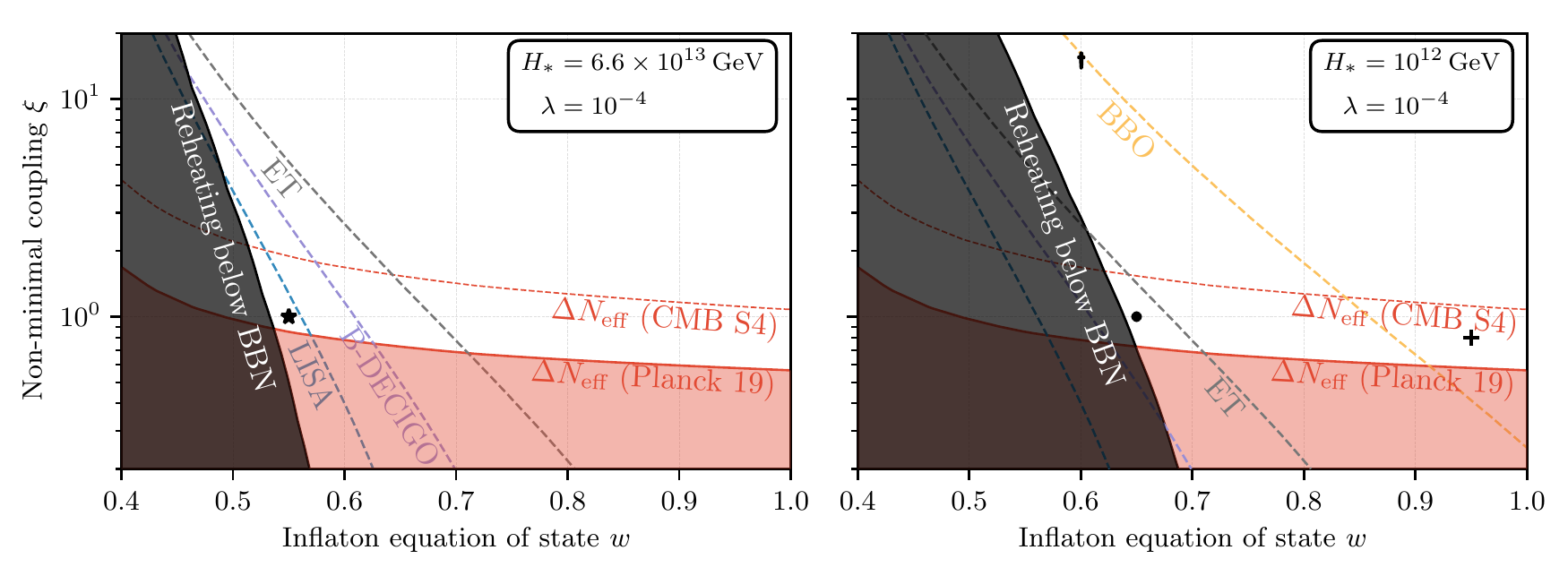}
	\caption{Detectability of the inflationary gravitational wave background in future experiments. In the left panel, the entire allowed parameter space is in reach of the BBO experiment. }
	\label{fig:moneyplot}
\end{figure*}

Below the scale $k_{\rm ke}^{0}$, the GW spectrum remains flat with an amplitude today given by~\cite{Caprini:2018mtu}
\begin{equation}
\Omega_{\rm GW}^{\rm 0, flat} = \frac{\Omega_{\gamma}^{0}}{24}  \left(\frac{g_{s,\rm eq}}{g_{s,k}}\right)^{\frac{4}{3}} \left(\frac{g_{k}}{g_{\gamma}^{0}} \right) \frac{2}{\pi^{2}} \frac{H_{*}^{2}}{M_P^{2}}\,,
\end{equation}
where $g_{\gamma}^{0} = 2$ and $g_{k}$ is the number of degrees of freedom when the scale $k$ crossed the horizon and ${\Omega_{\rm \gamma}^{0} = 5.38\times10^{-5}}$~\cite{Tanabashi:2018oca}.  We can now write the complete primordial GW spectrum today in piecewise fashion
\begin{equation}
	\Omega^{0}_{\rm GW} (k) = \Omega_{\rm GW}^{\rm 0, flat} \begin{cases}
		 1\,, & k < k_\text{ke}^{0} \\ 
		\left(\frac{k}{k_\text{ke}^{0}}\right)^{\frac{2(3w-1)}{1+3w}}\,, & k_\text{ke}^{0} \leq k \leq k_{*}^{0}\ \\
		0 & k > k_{*}^{0}
			\end{cases} \,.
	\label{eq:GWspecTot}
\end{equation}
In \cref{fig:GWspec}, we plot the spectra for the benchmark model parameters given in \cref{tab:BMPs} which can produce a detectable GW signal in future detectors such as LISA~\cite{Audley:2017drz}, ET~\cite{Sathyaprakash:2012jk}, B-DECIGO~\cite{Seto:2001qf,Sato:2017dkf}, and BBO~\cite{Crowder:2005nr}. 

To find the regions of parameter space which can be probed by these future GW detectors, we compute the signal-to-noise ratio (SNR) for each point using~\cite{Thrane:2013oya,Caprini:2015zlo}
\begin{align}
	\text{SNR}_\text{exp} \equiv \left\{ 2 t_\text{obs} \int_{f_\text{min}}^{f_\text{max}} d f \left[ \frac{h^2 \Omega_\text{GW}(f)}{h^2\Omega_\text{eff}(f)}\right]^2\right\}^{1/2}\,.
\end{align}
The observation time $t_\text{obs}$, frequency domain $f_\text{min/max}$ and noise curves $(h^2\Omega_\text{eff})$ for the different experiments are all taken from Table III in Ref~\cite{Breitbach:2018ddu}\footnote{Both the noise curves for the SNR as well as the power-law integrated sensitivity curves shown in \cref{fig:GWspec} are available in the ancillary material of Ref.~\cite{Breitbach:2018ddu}.}, in conjunction with the usual definition for $h$, namely $H_0=(100h)\SI{}{\km\per\s\per\mega\pc}$. The results are shown in the ${w-\xi}$ plane of \cref{fig:moneyplot} with fixed $\lambda = 10^{-4}$. The regions below the dashed lines have SNRs above the experimental threshold and are therefore testable. The black region is excluded by reheating temperatures below the onset of BBN ($T_\text{RH} \leq T_\text{BBN} \sim \SI{1}{\MeV}$), while regions violating $N_\text{eff}$ constraints, discussed below in \cref{sec:Neff}, are shaded red.

Assuming the detection of such a blue-tilted spectrum, the determination of the slope by one or more experiments would allow the equation of state $w$ of the kination phase to be probed and could thus confirm a non-standard period of cosmological history. While measuring $w$ would not confirm our model, it would strengthen the case for considering reheating scenarios of this type.

\section{Observational and Theoretical Constraints}
\label{sec:constraints}
\subsection{Overproduction of Gravitational Waves}
\label{sec:Neff}
The energy density of the blue-tilted primordial gravitational wave spectrum changes the number of effective relativistic degrees of freedom ($N_{\rm eff}$) at recombination time by an amount
\begin{equation}
\Delta N_{\rm eff} = \frac{8}{7} \left(\frac{11}{4} \right)^{\frac{4}{3}} \frac{\Omega_{\rm GW}^{0}}{\Omega_{\rm \gamma}^{0} }\,,
\end{equation}
where we have used the fact that the ratio $\Omega_{\rm GW}/\Omega_{\gamma}$ becomes constant after electron-positron annhilation at $T \sim 0.5$ MeV, so it is convenient to simply evaluate it at the present time instead of at recombination. The energy density in gravitational waves today is given by
\begin{equation}
\Omega_{\rm GW}^{0} =  \int \frac{df}{f} \Omega_{\rm GW}^{0}(f) \,.
\end{equation}
The Planck 2018 data requires $\Delta N_{\rm eff} < 0.284$ at 95\% C.L. using the TT,TE,EE+lowE+lensing+BAO dataset~\cite{Aghanim:2018eyx}, which is the value we use to set the $\Delta N_{\rm eff}$ exclusion line in our plots. Also shown in our plots is a $\Delta N_{\rm eff} < 0.03$ line, which is a conservative estimate of the sensitivity that the next generation of ground-based telescope experiments (CMB Stage-4) will be able to achieve~\cite{Abazajian:2016yjj}.

\subsection{Scalar Perturbations}
The power spectrum of superhorizon fluctuations of the reheaton field at the end of inflation for $\xi \gtrsim 0.1$ is given by
\begin{equation}
\mathcal{P}_{\delta\phi}(k) \approx  \left(\frac{H_{*}}{2\pi}\right)^{2} \left(\frac{H_{*}}{m_{*}}\right) \left(\frac{k}{H_{*}} \right)^{3} \,,
\label{eq:shPS}
\end{equation}
which is suppressed as $k^3$ for $k\ll H_{*}$ (see \cref{app:B} for a derivation). Thus, the reheaton fluctuations cannot account for the observed scale invariant spectrum of scalar perturbations. In order to achieve agreement with data, we must assume that the inflationary sector can produce the correct primordial curvature perturbations. However, since the superhorizon fluctuations of the reheaton field are power law suppressed, the curvature perturbation set by the inflationary sector will be adiabatic to a good approximation and will therefore be conserved on superhorizon scales. Thus, while the reheaton itself cannot produce the correct scalar perturbations, we can at least conclude that it will not interfere with them being set by the inflationary sector. This is consistent with findings that superhorizon curvature perturbations, once set, cannot be easily suppressed~\cite{Langlois:2004nn,Sloth:2005yx,Bartolo:2005jg}.

\section{Higgs as the Reheaton}
\label{sec:HiggsRH}
\begin{figure}
	\includegraphics[width=\columnwidth]{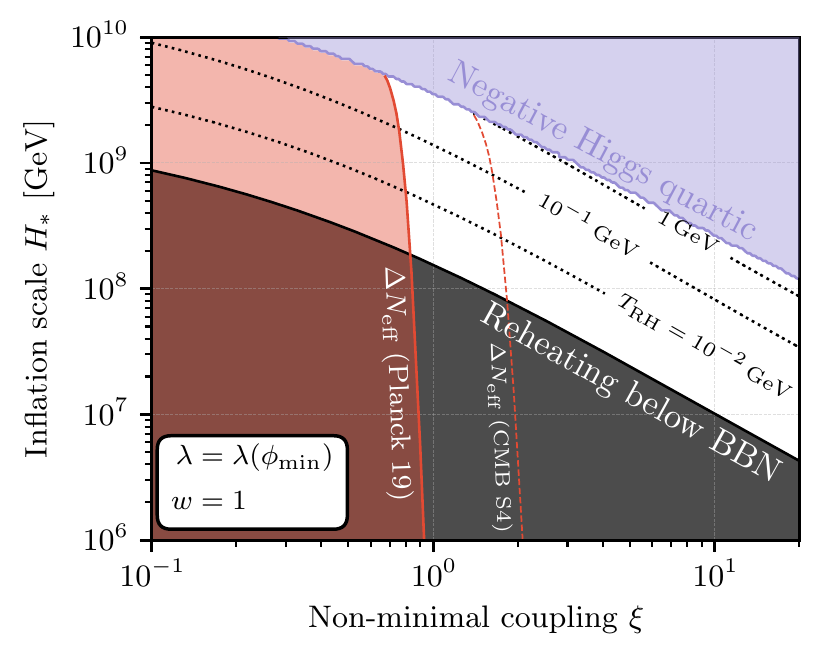}
	\caption{Reheating temperature contours in the case where the SM Higgs is identified as the reheaton field. The white region indicates the allowed parameter space. We take $w=1$ such that the reheating temperature is maximized.}
	\label{fig:HiggsRH}
\end{figure}

In order to consider the possibility of identifying the Higgs field as the reheaton, we need to know the size of the Higgs quartic at the energy scale of interest. To accomplish this, we assume the SM only running of Ref.~\cite{Buttazzo:2013uya} and renormalize the Higgs quartic at the scale $\mu = \phi_{\rm min}$. The consistency equation we must then solve for the location of the minimum is \cref{eq:phi_m} with $\phi_{\rm min} = \mu$ and $\lambda = \lambda(\mu)$
\begin{equation}
\mu^{2} = \frac{3\xi(3w-1)}{\lambda(\mu)} H^{2}_{\rm min}(\lambda(\mu))\,.
\end{equation}
Solutions to this equation exist only while $\lambda(\phi_{\rm min}) > 0$, which places an upper bound on $H_{*}$ for fixed $\xi$ since the SM Higgs quartic runs negative at high energies. We show this bound in the $\xi$-$H_{*}$ plane of \cref{fig:HiggsRH}, where we also plot contours of constant reheating temperatures assuming that $T_{\rm RH}$ is given by the $\Gamma > H_{\rm matt}$ case of \cref{sec:mainRH}. This assumes the decay of the Higgs and equilibration of the resulting SM radiation happens before reheaton-kination equality, which is expected to be the case~\cite{Enqvist:2013kaa,Figueroa:2014aya,Enqvist:2014tta,Figueroa:2015rqa,Enqvist:2015sua,Figueroa:2016dsc}. For fixed $\xi$, we find that an allowed range of inflaton scales $H_{*}$ exist where reheating before BBN can be achieved without field excursions which probe the unstable region of the SM Higgs potential. For $\xi \sim 1$, the allowed range is $H_{*} \sim 10^{8} - 5\times 10^{9}$ GeV with reheating temperatures $T_{\rm RH} \sim 10^{-3} -1$ GeV.

We note that the idea of reheating using a non-minimally coupled Higgs was first investigated by Ref.~\cite{Figueroa:2016dsc}.  In that work, however, the time variation of the Hubble rate as the reheaton rolls to the minimum was neglected. As we showed in \cref{section:TttM}, we find that the reheaton typically takes $\mathcal{O}(1)$ e-fold to roll to the minimum, by which time the Hubble rate has typically decreased by a factor $H_{\rm min} \sim 10^{-2} H_{*}$. Since the energy extracted scales as $H_{\rm min}^{4}$, neglecting the time variation of $H$ results in an overestimation of the energy density by a factor of $10^{8}$ for typical values of the parameters.

\section{Conclusions}
\label{sec:conc}

While there is now compelling evidence for a period of exponential inflation in the early universe, much less is known about reheating. This process describes the transition from a quasi de-Sitter universe dominated by the inflaton energy density to a thermal plasma dominated by SM particles that is at least hot enough to allow for successful nucleosynthesis. A particularly interesting question is whether this transition is possible in the absence of any non-gravitational couplings of the inflaton to the SM.

Recently it was argued that gravitational reheating is not viable due to an overproduction of gravitational radiation~\cite{Figueroa:2018twl}. Here we show both analytically and numerically that these constraints can be evaded if a scalar field, the so called {\it reheaton}, has a non-minimal coupling to gravity and a period of kination occurs after the end of inflation. The resulting tachyonic instability in the reheaton allows it to more efficiently extract energy from the gravitational background, reducing the relative importance of the gravitational radiation. Decays of the reheaton to SM particles can then populate the SM thermal bath while the inflaton energy dilutes away faster than radiation. This permits reheating temperatures which are high enough to allow for successful thermal leptogenesis.

The stochastic GW background which previously cast doubt on the viability of gravitational reheating now becomes the main hope for probing this reheating scenario. The blue-tilted GW spectrum is in range of several future GW detection experiments, at least for inflation scales close to the maximally allowed value. In addition, future CMB observations will improve the bound on $\Delta N_{\rm eff}$ and thus further constrain the high frequency end point of the spectrum. 

Finally, we have explored the intriguing possibility that the Higgs field itself can play the role of the reheaton. In this scenario, no additional fields beyond the SM would be required. We extend previous studies of this scenario by taking into account the variation of the Hubble rate during the initial rolling of the reheaton, resulting in strongly suppressed reheating temperatures. Nevertheless, a small window with reheating temperatures up to a GeV remains viable. 


\section*{Acknowledgements}
We thank Valerie Domcke, Dani Figueroa, and Alberto Salvio for useful discussions.
Research in Mainz is supported by the Cluster of Excellence ``Precision Physics, Fundamental Interactions, and Structure of Matter" (PRISMA+ EXC 2118/1) funded by the German Research Foundation (DFG) within the German Excellence Strategy (Project ID 39083149). TO is supported by the European Research Council (ERC) under the European Union's Horizon 2020 research and innovation programme (grant agreement No. 637506 ``$\nu$Directions'').

\appendix

\section{Energy-momentum of a non-minimally coupled scalar field}
\label{app:A}
Varying the action in \cref{eq:act} with respect to $g^{\mu\nu}$ yields the Einstein equations
\begin{equation}
G_{\mu\nu} = R_{\mu\nu} - \frac{1}{2} g_{\mu\nu} R = \frac{1}{M_P^{2}}\left( T_{\mu\nu}^{\phi} + T_{\mu\nu}^{\rm m}\right) \,,
\end{equation}
where the energy-momentum tensor $T^{\phi}_{\mu\nu} $ of the non-minimally coupled scalar field $\phi$ is~\cite{Ford:1987de,Ford:2000xg,PhysRevD.11.2072,PhysRevD.54.6233,DESER1984419,CALLAN197042,Bellucci:2001cc,Hrycyna:2015vvs}
\begin{equation}
\begin{split}
T^{\phi}_{\mu\nu} &= \partial_{\mu}\phi\partial_{\nu}\phi - g_{\mu\nu}\left( \frac{1}{2}g^{\rho\sigma}\partial_{\rho}\phi\partial_{\sigma}\phi - V(\phi)\right) \\
&+ \xi (G_{\mu\nu} + g_{\mu\nu} \nabla^{\sigma}\nabla_{\sigma} - \nabla_{\mu}\nabla_{\nu}) \phi^{2} \,.
\label{eq:TmunuPhi}
\end{split}
\end{equation}
For the metric in \cref{eq:ds2} and a homogeneous field $\phi = \phi(t)$, one can show that \cref{eq:TmunuPhi} reduces to the form of a perfect fluid with energy density
\begin{equation}
\rho_{\phi} = \frac{1}{2}\dot{\phi}^{2} + V(\phi) + \xi\left(3H^{2}\phi^{2} + 6 H \phi \dot{\phi}\right)  \,,
\end{equation}
and pressure
\begin{equation}
p_{\phi}  = \frac{1}{2}\dot{\phi}^{2} - V(\phi) + \frac{\xi}{3} (R + 3H^{2} )\phi^{2} -2\xi \left( \dot{\phi}^{2} + \phi \ddot{\phi}+ 2H\phi\dot{\phi} \right)\,.
\end{equation}
Using the equation of motion for $\phi$, we have $\ddot{\phi} = \xi R \phi - 3H\dot{\phi} - V'(\phi)$ which combined with \cref{eq:ricci} allows $p_{\phi}$ to be written as
\begin{equation}
\begin{split}
p_{\phi} &= \frac{1}{2}\dot{\phi}^{2} - V(\phi) + 3\xi \left[ w+2\xi(1-3w)\right]H^{2}\phi^{2} \\
&+ 2\xi\left(V'\phi  +H\phi\dot{\phi}   -\dot{\phi}^{2}\right) \,.
\end{split}
\end{equation}
We now define the equation of state $w_{\phi}$ of the non-minimally coupled reheaton as $w_{\phi} = p_{\phi} / \rho_{\phi}$. Of particular interest is the initial reheaton equation of state during the initial tachyonic growth phase while the reheaton is rolling to the minimum. In this phase, the bare potential $V$ is negligible, so the initial equation of state can be computed as
\begin{equation}
\begin{split}
&w^{\rm init}_{\phi} = p_{\phi}^{\rm init} / \rho_{\phi}^{\rm init}  \\
&=\frac{\frac{1}{2}\dot{\phi}^{2}+ 3\xi \left[ w+2\xi(1-3w)\right]H^{2}\phi^{2}  + 2\xi\left( H\phi \dot{\phi} -\dot{\phi}^{2} \right)}{\frac{1}{2}\dot{\phi}^{2} + \xi(3 H^{2}\phi^{2}+ 6H \phi \dot{\phi} )} \,.
\label{eq:init_EoS}
\end{split}
\end{equation}
During the tachyonic growth phase, the time evolution of $\phi$ is given by \cref{eq:phi_scal}, from which we can derive that $\dot{\phi} = -\gamma \phi \dot{H}/H = 3\gamma(1+w) H \phi /2$. Using this relation to eliminate $\dot{\phi}$ in \cref{eq:init_EoS} causes $\phi$ and $H$ to drop out and we can write the initial reheaton equation of state entirely in terms of the inflaton equation of state $w$ and the non-minimal coupling $\xi$
\begin{equation}
w^{\rm init}_{\phi} = \frac{8\xi[\gamma+w(1+\gamma-6\xi)+2\xi] -3\gamma^{2}(4\xi -1)(1+w)^{2}}{8\xi+3\gamma(1+w)(\gamma+w\gamma+8\xi)} \,,
\label{eq:init_EoS_2}
\end{equation}
with $\gamma = \sqrt{\xi(3w-1)/3}$, the same definition used in \cref{eq:phi_scal}. The scaling of the reheaton energy density in the initial tachyonic phase can now be written in terms of its equation of state
\begin{equation}
\rho_{\phi}^{\rm init} \propto a^{-3(1+w^{\rm init}_{\phi})} \,,
\end{equation}
which allows us to analytically determine the initial slope of the lines in \cref{fig:phi_energy_density}.
While \cref{eq:init_EoS_2} is a rather unwieldy expression, it has some insightful limits. For $w=1/3$, we have $\gamma = 0$ and thus $w^{\rm init}_{\phi} = 1/3$. As expected, in this limit there is no tachyonic growth and the reheaton energy simply behaves as radiation. In the opposite limit $w=1$, we find
\begin{equation}
w^{\rm init}_{\phi} (w=1) = 1-\sqrt{\frac{8\xi}{3}} \,.
\end{equation}
We see that there is a critical value of $\xi = 3/2$ where $w^{\rm init}_{\phi} = -1$ and the reheaton energy behaves initially as a cosmological constant. The interpretation here is that the tachyonic growth is exactly compensating the dilution of the reheaton energy due to the expansion of the universe. For $\xi > 3/2$, the tachyonic growth overcomes the expansion, resulting in a net growth of the reheaton energy density. For $1/6 < \xi < 3/2$, the reheaton energy decays, but it does so at a rate slower than radiation. This analysis explains why the $\xi = 5,10$ lines in \cref{fig:phi_energy_density} have a positive (growing) initial slope whereas the $\xi = 1$ line has a negative (decaying) initial slope which is less steep than the gravitational particle production line (which scales as radiation).

For general $w$ in the range $1/3 < w \leq 1$, there exists a critical value of $\xi$ which we denote as $\xi_{\rm crit}$ where the reheaton energy scales initially as a cosmological constant. This value is given by solving \cref{eq:init_EoS_2} for $\xi$ with $w^{\rm init}_{\phi} = -1$, but for $w \gtrsim 0.5$, it can be reasonably approximated by the condition found in \cref{sec:init_dyn} such that the reheaton rolls to the minimum faster than the minimum approaches the origin
\begin{equation}
\xi_{\rm crit} \approx \frac{1}{w-1/3} \,,
\end{equation}
and the approximation becomes exact in the case $w=1$. Roughly speaking, the significance of $\xi_{\rm crit}$ is that for $\xi > \xi_{\rm crit}$, we expect the initial growth of the reheaton energy density to be faster than the dilution due to the expansion of the universe.

\section{Initial Conditions from de-Sitter}
\label{app:B}
To study the quantum fluctuations of the reheaton field during inflation, we work with the conformal time $dt = a\, d\tau$ where the field can be canonically normalized via the field redefinition $\phi \rightarrow \varphi /a$. The canonically normalized action is then
 \begin{equation}
\mathcal{S}_{\varphi} = \frac{1}{2} \int d\tau d^{3}x \left[(\varphi')^{2} - (\nabla\varphi)^{2} + a^{2}\left(\xi - \frac{1}{6}\right) R \varphi^{2} \right] \,,
\end{equation}
which leads to the following equation of motion for $\varphi$ in momentum space
 \begin{equation}
\varphi''_{k} + \left[k^{2} - a^{2}\left(\xi - \frac{1}{6}\right) R \right] \, \varphi_{k} = 0 \,.
\label{eq:mode_eq}
\end{equation}
In de-Sitter space, we have $R(\tau) = -12H^{2}(\tau)= -12 /(a\tau)^{2}$ and \cref{eq:mode_eq} can be cast in the form of Bessel's equation
 \begin{equation}
\varphi''_{k} + \left[k^{2} - \frac{\nu^{2} - 1/4}{\tau^{2}} \right] \, \varphi_{k} = 0 \,,
\label{eq:bessel_eq}
\end{equation}
with
\begin{equation}
\nu^{2} \equiv \frac{1}{4} - 12\left(\xi - \frac{1}{6}\right)\,.
\end{equation}
The general solution of \cref{eq:bessel_eq} can be written as
\begin{equation}
\varphi_{k}(\tau) = \sqrt{-\tau}\left[ \alpha_{k}  H_{\nu}^{(1)}(-k\tau) + \beta_{k}  H_{\nu}^{(2)}(-k\tau) \right]\,.\,
\label{eq:gen_sol}
\end{equation}
where $H_\nu^{(1)}(x)$ and $H_\nu^{(2)}(x)$ are Hankel functions of the first and second kind, respectively.

\subsection{Canonical quantization}
We canonically quantize following standard methods of quantum field theory in curved spacetime~\cite{Birrell:1982ix,Mukhanov:2007zz}. First, we promote $\varphi_{k}$ to an operator
\begin{equation}
\hat{\varphi}_{k}(\tau) = \varphi_{k}(\tau) \hat{a}_{k} + \varphi_{k}^{*} (\tau) a^{\dagger}_{-k} \,,
\end{equation}
where $[\hat{a}_{k},\hat{a}_{k'}^{\dagger}] = (2\pi)^{3} \delta(k+k')$ and the modes are normalized such that $\varphi_{k}\varphi_{k}^{\prime *} -\varphi_{k}' \varphi_{k}^{*} = i$. To find the coefficients of \cref{eq:gen_sol} one constructs the Hamiltonian operator and minimizes the vacuum expectation value of the energy $\langle 0 | \hat{H} | 0 \rangle$. Assuming that all modes are well within the horizon at early times, this leads to the requirement that $\varphi_{k}$ should approach the Bunch-Davies vacuum for $k\tau \rightarrow -\infty$
\begin{equation}
\varphi_{k} (k\tau \rightarrow -\infty) \approx \frac{1}{\sqrt{2k}} e^{-ik\tau} \,,
\end{equation}
which one can show using the asymptotic expansion for the Hankel functions requires $\beta_{k} = 0$ and $\alpha_{k} = \sqrt{\pi/4} \, e^{ i\frac{\pi}{4}(1+2\nu)}$. The solution for $\varphi_{k}$ in de-Sitter space is then
\begin{equation}
\varphi_{k}(\tau) = \sqrt{\frac{\pi}{4k}}  e^{ i\frac{\pi}{4}(1+2\nu)} \sqrt{- k\tau} \, H_{\nu}^{(1)}(-k\tau) \,.
\label{eq:dssol}
\end{equation}
The operator in position space can be found by Fourier transforming the field
\begin{equation}
\hat{\varphi} (\tau, x) = \int \frac{d^{3} k}{(2\pi)^{3}} \left[ \varphi_{k}(\tau) \hat{a}_{k} + \varphi_{k}^{*} (\tau) a^{\dagger}_{-k}\right] e^{ikx} \,,
\end{equation}
and expectation values of the original field $\phi = \varphi / a$ such as the two-point function can be computed as
\begin{equation}
\langle \hat{\phi} ^{2} \rangle = \langle 0| \hat{\phi}^{*} (\tau, 0)\hat{\phi} (\tau, 0) |0 \rangle = \int \frac{dk}{k} \mathcal P_{\phi}(k,\tau) \,,
\end{equation}
with 
\begin{equation}
\mathcal{P}_{\phi}(k,\tau) = \frac{k^{3}}{2\pi^{2}a^{2}} |\varphi_{k}(\tau)|^{2} .
\label{eq:PS}
\end{equation}
Similarly, the power spectrum of $\phi' = d\phi/d\tau$ can be computed as
\begin{equation}
\begin{split}
&\mathcal{P}_{\phi'}(k,\tau) = \frac{k^{3}}{2\pi^{2}} \bigg|\frac{d}{d\tau}\left(\frac{\varphi_{k}}{a}\right)\bigg|^{2} 
=  \frac{k^{3}}{2\pi^{2}a^{2}} |\varphi'_{k}-aH\varphi_{k}|^{2} \,.
\end{split}
\label{eq:velvar} 
\end{equation}

\subsection{Variances at the end of inflation}
For $\xi > 3/16$, we are in the regime where the reheaton is heavy during inflation with an effective mass $m_{\rm eff}^{2} = 12 (\xi-1/6)H^{2}  \gtrsim H^{2}$. In this case, $\nu$ is purely imaginary and the power spectrum of $\phi$ for $\nu =i\mu$ is
\begin{equation}
\mathcal{P}_{\phi}(k,\tau) = \frac{H^{2}}{8\pi}(-k\tau)^{3} |H_{i\mu}^{(1)}(-k\tau)|^{2} e^{ -\pi \mu} \,,
\label{eq:imPS}
\end{equation}
with
\begin{equation}
\mu^{2} \equiv 12\left(\xi - \frac{1}{6}\right) - \frac{1}{4}\,.
\end{equation}
We now work in the variable $z= -k\tau = k/(aH) $, where the superhorizon modes are those with $z<1$.
For $\mu \gg 1$ and in the superhorizon limit, one can show using the small argument expansion for the Hankel functions that $|H_{i\mu}^{(1)}(z)|^{2} \, e^{ -\pi \mu} \approx 2/(\pi\mu)$. This leads to a superhorizon power spectrum for $\phi$ at the end of inflation of
\begin{equation}
\mathcal{P}_{\phi}(z,\tau_{*}) = \frac{H_{*}^{2}}{4\pi^{2}}\frac{z^{3}}{\mu} \approx  \frac{H_{*}^{2}}{4\pi^{2}}\frac{z^{3}}{\sqrt{12\xi}} \,.
\end{equation}
In the same limit, one can show that the velocity power spectrum is
\begin{equation}
\mathcal{P}_{\phi'}(z,\tau_{*}) = \mu^{2}H^{2}_{*} \,\mathcal{P}_{\phi}(z,\tau_{*}) \approx \frac{H_{*}^{4}}{4\pi^{2}} z^{3} \sqrt{12\xi} \,.
\end{equation}
Since the superhorizon modes are those with $z < 1$, these power spectra are peaked at the horizon at the end of inflation and power law suppressed ($\propto z^3$) for modes well outside the horizon. To estimate the classical field value at the end of inflation, we integrate over the superhorizon fluctuations $z\lesssim 1$  to obtain
\begin{equation}
\langle \hat{\phi^{2}_{*}} \rangle =  \int_{0}^{1} \frac{dz}{z} \mathcal P_{\phi}(z,\tau_{*}) \approx \left( \frac{H_{*}}{2\pi} \right)^{2} \frac{1}{3\sqrt{12\xi}} \,,
\end{equation}
\begin{equation}
\langle \hat{\phi}'^{2}_{*} \rangle =  \int_{0}^{1} \frac{dz}{z} \mathcal P_{\phi'}(z,\tau_{*}) \approx \frac{H^{4}_{*}}{12\pi^{2}} \sqrt{12\xi} \,.
\end{equation}
This is the same result as can be found in Ref.~\cite{Cosme:2018nly}. We acknowledge that a more detailed study of excitation of the field during the transition could find larger initial fluctuations, however Ref.~\cite{Herranen:2015ima} found a similar amplitude for a transition in which $R$ does not change sign. 

\subsection{Initial Conditions}
\begin{figure}
	\includegraphics[width=\columnwidth]{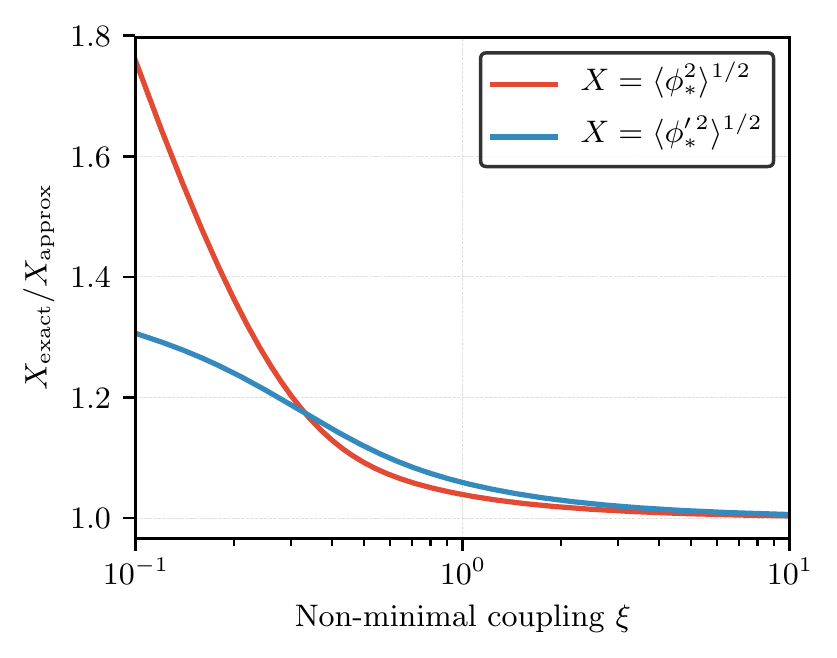}
	\caption{Ratios of the exact numerical solution for the RMS field value to the approximations given in \cref{eq:RMSphi,eq:RMSphidot}.}
	\label{fig:IC_comp}
\end{figure}

We take the root mean square (RMS) fluctuations of the reheaton at the end of inflation as our initial conditions for the classical field evolution. Defining $m_{*} = \sqrt{12\xi}H_{*}$ yields
\begin{equation}
\phi_{*}  = \sqrt{\langle \hat{\phi^{2}_{*}} \rangle} \approx \frac{H_{*}}{2\pi} \sqrt{\frac{H_{*}}{3m_{*}}} \,, 
\label{eq:RMSphi}
\end{equation}
\begin{equation}
 \dot{\phi}_{*} = \sqrt{\langle \hat{\phi}'^{2}_{*} \rangle}\approx m_{*} \phi_{*} \,,
 \label{eq:RMSphidot}
\end{equation}
which are the initial conditions given in \cref{eq:phi_IC}. We note that these expressions were derived under the assumption that $\mu \gg 1$, which translates into $\xi \gtrsim 1$. However, it is a reasonable approximation to extend their range of validity down to $\xi \gtrsim 0.1$. We justify this statement by plotting in \cref{fig:IC_comp} the ratio of the exact numerical solution for the RMS field value found by inserting \cref{eq:dssol} into \cref{eq:PS} and \cref{eq:velvar} to the approximations given in \cref{eq:RMSphi,eq:RMSphidot}. We see good agreement for $\xi \gtrsim 1$ and a difference of $\lesssim 2$ for $\xi \sim 0.1$.

\section{Dynamics of a non-minimally coupled scalar}
\label{app:NMC_dyn}

\begin{figure*}
\begin{center}
	\includegraphics[width=0.495\textwidth]{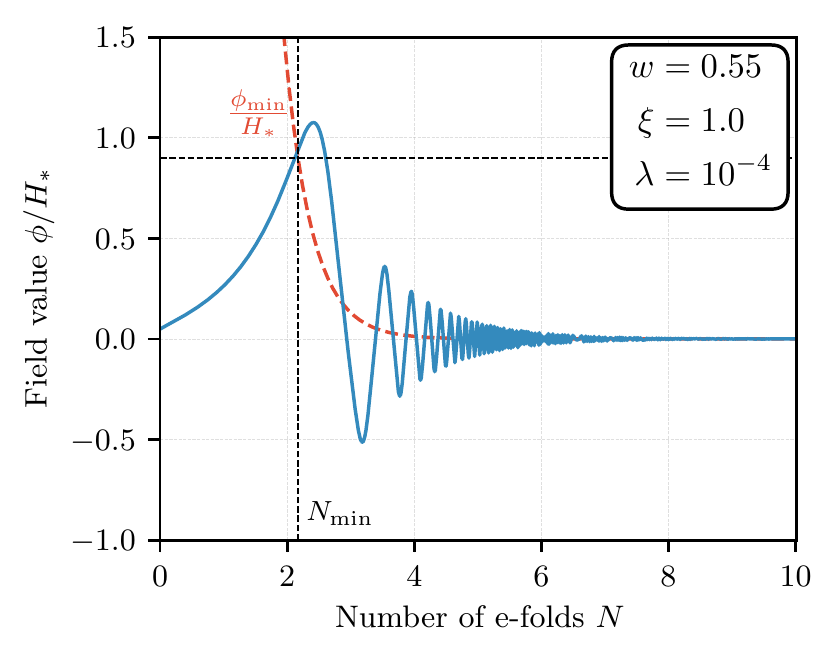}
	\includegraphics[width=0.495\textwidth]{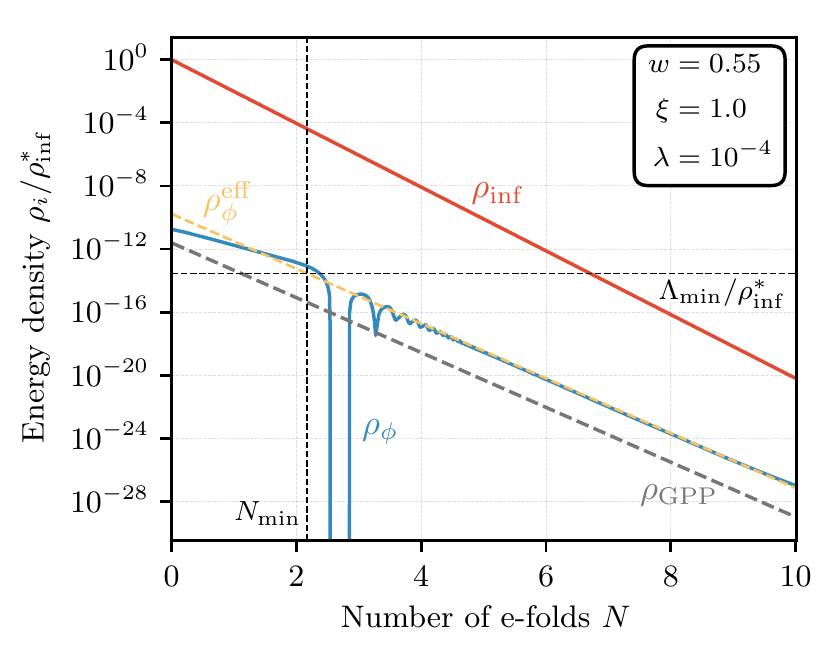}
	\caption{Dynamics of the reheaton directly after inflation ends for the LISA benchmark point 1 from \cref{tab:BMPs}. In the left-hand panel we show the field displacement as a function of time (blue) as well as the location of the minimum (red-dashed). In the right-hand panel the energy density is shown for the reheaton (blue), inflaton (red) as well as the standard gravitational particle production energy density $\rho_\text{GPP} \sim \SI{E-2}{} H_*^4$ (grey-dashed). In both panels the approximations from the main body of the paper are shown as black dashed lines, while the effective energy density calculated using these quantities is shown in the right-hand panel (yellow-dashed). }
	\label{fig:dynamics}
\end{center}
\end{figure*}

To support the analytic approximations given throughout the text, we now turn to numerically solving the full differential equation describing the time evolution of the classical field. In \cref{fig:dynamics} we show the results of solving \cref{eq:eom} using the initial conditions in \cref{eq:phi_IC} as a function of the number of e-folds. The left-hand panel shows the field value, while the right-hand side shows the energy density calculated using \cref{eq:rho_phi}. In both panels the dashed lines show the approximations described in the text for $N_\text{min}$, $\phi_\text{min}$ and $\Lambda$. We find good agreement between the $\rho_\phi^\text{eff}$ approximation and the full numerical result at late times. Other parameter choices which better illustrate the initial tachyonic growth as well as a period of matter scaling are shown in \cref{fig:phi_energy_density} along with their respective $\rho_\phi^\text{eff}$ as dashed lines.

\section{Details of the gravitational wave stochastic background}
\label{app:GW}

The flat GW inflationary power spectrum as each mode $k$ re-enters the horizon is given approximately by \cite{Boyle:2005se,Caprini:2018mtu}
\begin{align}
	\mathcal{P}(k) \simeq \frac{2}{\pi}\frac{H_*^2}{M_P^2}\,,
\end{align}
where the resulting energy density today normalized to the critical energy density takes the form
\begin{align}
	\Omega^0_\text{GW}(k) &\simeq \frac{k^2}{12 H_0^2 a_0^2} T^2(k,k_0)\mathcal{P}(k)\,.
\end{align}
$T^2(k,k_0)$ is the {\it transfer function} which encodes the red- or blue-shifting of the mode functions as well as their evolution as the scaling of the dominant component of the energy density changes. In what follows we neglect the matching of the mode functions at both the transition between kination and radiation (and or matter domination) and the standard transition between radiation and matter in the late universe. The later occurs at frequencies much below the experiments of interest, while the $H_* \lesssim \SI{6.6E+13}{\GeV}$ constraint accounts for CMB constraints on the tensor spectrum \cite{Akrami:2018odb}. As a result the transfer function takes the simple form 
\begin{align}
T^2(k,k_0) &= \frac{1}{2} \left(\frac{a_k}{a_0}\right)^2\,.
\end{align}
As a result we obtain
\begin{align}
	\Omega^0_\text{GW}(k) &\simeq \frac{1}{24} \left(\frac{H_k}{H_0}\right)^2\left(\frac{a_k}{a_0}\right)^4 \mathcal{P}(k)\,,
\end{align}
where we have used $k = a_k H_k$ for the comoving momenta at the scale where the mode re-enters the horizon. Note that the amount of blue and/or red-shifting depends upon whether $k \geq k_\text{ke}$. Before we derive this expression we first need $H_\text{ke}$.

Using the effective energy density, c.f. \cref{rho_phi_eff}, kination ends when $\rho_\phi^\text{eff}$ or equivalently the energy density of its decay products are equal to the inflaton energy density
\begin{align}
	\rho_{\phi,*}^\text{eff} \left(\frac{a_*}{a_{\rm ke}} \right)^{4} = 3M_{P}^{2}H_{*}^{2}\left(\frac{a_*}{a_{\rm ke}} \right)^{3(1+w)} \,.
\end{align} 
In this expression, we assume the effective energy density scales as radiation. We show how a period of matter scaling would modify this expression below. At the end of kination, the Hubble rate is given by
\begin{align} \label{eq:HubbleAtHke}
	H_{\rm ke}^2 =\frac{2}{3M_P^2} \rho_{\rm inf, ke} = 2H_*^{2}\left(\frac{a_*}{a_{\rm ke}} \right)^{3(1+w)} \,,
\end{align}   
and eliminating the ratio of scale factors yields
\begin{align}\label{eq:HkeApp}
	\frac{H_{\rm ke}}{\sqrt{2} H_*} = \left(\frac{\rho_{\phi,*}^\text{eff}}{3M_{P}^{2}H_*^{2}}\right)^{\frac{3(1+w)}{2(3w-1)}} \equiv \Omega_{\phi,*}^{\frac{3(1+w)}{2(3w-1)}} \,.
\end{align}  
In the scenario where the bare mass of the reheaton becomes relevant before $H_\text{ke}$, i.e. $\Gamma_\text{matt} > H_\text{ke}$, a period of matter scaling reduces the redshifting of the reheaton energy density. As $\rho_\phi^\text{eff}$ scales as radiation both before and after this period of matter scaling, we need only multiply $\rho_{\phi,*}^\text{eff}$ by a factor that compensates for this decreased dilution, namely 
\begin{align}
	\frac{a_\text{matt}}{a_\text{decay}} &= \left(\frac{H_\text{matt}}{\Gamma}\right)^{\frac{2}{3(1+w)}}\,,
\end{align} 
where $a_\text{decay}$ is the scale factor when $\Gamma = H$. Including this factor yields the full expression in \cref{eq:Hke}.

With $H_\text{ke}$ in hand we can now compute the physical momenta of the horizon size redshifted to today at both kination-reheaton equality and the end of inflation. The former is given already in \cref{eq:kke0}, while the latter is
\begin{align}
	k_*^0 &= H_* \left(\frac{a_*}{a_\text{ke}} \right)\left(\frac{a_\text{ke}}{a_0}\right)\,, \notag \\
	 &=H_{*} \left(\frac{g_{s, \rm eq} }{g_{s,\rm ke}} \right)^{\frac{1}{3}} \left(\frac{T_{0}}{T_{\rm ke}} \right) \Omega_{\phi,*}^{\frac{1}{3w-1}} \,,
\end{align}
where we have used \cref{eq:HubbleAtHke,eq:HkeApp} in the second line. An important feature realized in Ref.~\cite{Figueroa:2018twl} is that $k_*^0$ is independent of Hubble at the end of inflation. This can be seen by inserting \cref{eq:TrhMAIN}, where again $T_\text{RH}=T_\text{ke}$, in conjunction with the observation that $\Omega_{\phi,*} \propto H_*^2$:
\begin{align}
	k_*^0 &= T_0 \left(\frac{g_{s, \rm eq} }{g_{s,\rm ke}} \right)^{\frac{1}{3}} \left(\frac{\pi^2 g_{\text{RH}}}{90} \frac{H_*^2}{M_P^2}\right)^{\frac{1}{4}}  \Omega_{\phi,*}^{-\frac{1}{4}}\,.
	\label{eq:kstar0}
\end{align}
Consequently the $N_\text{eff}$ constraint, which depends largely on the GW amplitude at the scale $k_*^0$, is independent of the scale of inflation as can be seen in \cref{fig:genericRH}. This statement is no longer true if there was a period where the reheaton scaled as matter, in which case \cref{eq:kstar0} should be multiplied by a factor $(H_{\rm matt} /\Gamma)^{-1/(6+6w)}$, introducing a weak dependence on $H_{*}$.

Finally, we return to the GW energy density, which for $k > k_\text{ke}$ is
\begin{align}
	\Omega^0_\text{GW}(k) &\simeq \frac{1}{24} \left(\frac{H_k}{H_\text{ke}}\frac{H_\text{ke}}{H_0}\right)^2\left(\frac{a_k}{a_\text{ke}}\frac{a_\text{ke}}{a_0}\right)^4 \mathcal{P}(k)\,, \notag \\
	&= \frac{1}{24} \left(\frac{a_k}{a_{\rm ke}} \right)^{1-3w} \left(\frac{H_\text{ke}}{H_0}\right)^2\left(\frac{a_\text{ke}}{a_0}\right)^4 \mathcal{P}(k)\,, \notag \\
	&= \left(\frac{k}{k_\text{ke}^0}\right)^{\frac{2(3w-1)}{1+3w}} \Omega_{\rm GW}^{\rm 0, flat} \,,
\end{align}
where we have made the sudden transition approximation where $H_{\rm ke}^{2}/H_{k}^{2} = (a_k / a_{\rm ke})^{3(1+w)}$. This leads to the piecewise result in \cref{eq:GWspecTot}. 
\bibliographystyle{JHEP}

\providecommand{\href}[2]{#2}\begingroup\raggedright\endgroup

\end{document}